\documentclass[a4paper,11pt]{article}
\usepackage{pos}
\usepackage{mciteplus}

\newcommand{\jpsi}{$\mathrm{J}/\psi$ }
\newcommand{\jpsim}{\mathrm{J}/\psi}

\title{Electron-Ion Collisions at the LHeC and FCC-he}
\manuallySeparateAuthors
\author*{Heikki Mäntysaari }
  \author{for the LHeC and FCC-he Study Group}
\affiliation{
Department of Physics, University of Jyväskylä,  P.O. Box 35, 40014 University of Jyväskylä, Finland
}
\affiliation{
Helsinki Institute of Physics, P.O. Box 64, 00014 University of Helsinki, Finland
}
\emailAdd{heikki.mantysaari@jyu.fi}
\renewcommand{\printHeadAuthors}{Heikki Mäntysaari}

\abstract{
The LHeC and the FCC-he will open a new realm in our understanding of nuclear structure and the dynamics in processes involving nuclei, in an unexplored kinematic domain. We review some of the recent studies as shown in the update of the 2012 LHeC CDR, including the determination of nuclear parton densities in the framework of global fits and for a single nucleus,  inclusive and exclusive diffraction and the unique capabilies of these high-energy colliders for probing QCD in the non-linear regime of phase space.
}

\FullConference{%
  40th International Conference on High Energy physics - ICHEP2020\\
  July 28 - August 6, 2020\\
  Prague, Czech Republic (virtual meeting)
}

\begin{document}
\maketitle

\section{Introduction}

Thanks to their pointlike structure, electron beams can be used to accurately probe the internal structure of hadrons in Deep Inelastic Scattering (DIS) experiments. In electron-proton collisions, this is demonstrated by a precise extraction of the partonic structure of protons at HERA down to  small longitudinal momentum fraction $x$, where a rapid increase of parton densities have been observed~\cite{Abramowicz:2015mha}.

The observed growth of the parton densities can not continue down to asymptotically small values of $x$, as eventually the self-interactions between the gluons start to dominate, and one enters in the non-linear regime of the ``cold QCD phase diagram''.  These self interactions dynamically generate a scale, referred as the saturation scale $Q_s^2$, which is in the perturbative domain when densities are high enough. To reach these high densities, very large center-of-mass energies are required. An additional enhancement of parton densities is obtained by replacing protons by heavy nuclei, as the local parton densities scale as $A^{1/3}$, where $A$ is the nuclear mass number.

In order to experimentally access the non-linear  QCD dynamics at the highest possible densities, new high-energy and high-luminosity electron-ion collider facilities have been proposed at CERN. The two options covered in the recent update of the Conceptual Design Report (CDR)~\cite{Agostini:2020fmq} are the LHeC, where electrons would be collided with the protons accelerated at the LHC, and FCC-he. where electron-proton collisions would be studied in the Future Circular Collider~\cite{Abada:2019lih}. In addition, there are similar but lower center-of-mass energy plans  in the US~\cite{Accardi:2012qut,Aschenauer:2017jsk} and in China~\cite{Chen:2018wyz}.

The LHeC/FCC-he has a broad physics program including, for example, studies related to the Higgs boson, physics beyond the standard model and so on, see Ref.~\cite{Agostini:2020fmq}. In this Talk, I review some of the most important aspects of the LHeC/FCC-he physics program related to the QCD dynamics in electron-ion collisions.

\section{Nuclear parton distribution functions}
The electron-proton collisions measured at HERA have enabled precise extractions of the quark and gluon distribution functions for the proton~\cite{Abramowicz:2015mha} over a wide kinematical range, down to $x\sim 10^{-4}$ in perturbative scales. In case of nuclear parton distribution functions (nPDFs), the situation is very different due to the lack of high energy nuclear DIS data, resulting in large uncertainties in the nPDFs at small $x$. Recent nPDF extractions have included data from the proton-lead collisions measured at the LHC sensitive to small-$x$ structure, but the LHC data has only a limited impact on the nPDF extractions~\cite{Eskola:2016oht}.

To determine the LHeC impact on the nuclear PDF extractions, a modified version of the EPPS16~\cite{Eskola:2016oht} nuclear PDF parametrization, referred as EPPS16$^*$ has been fitted to the LHeC pseudodata. The modification is required in the impact analysis, as the original EPPS16 parametrization includes (physically motivated) assumptions that are used to extrapolate the nPDFs at small $x$ outside the region constrained by the data. These assumptions are relaxed in EPPS16$^*$, and consequently the uncertainty bands more accurately reflect how accurately the data constraints the small-$x$ structure of nuclei.

\begin{figure*}[tb]
\includegraphics[width=\textwidth]{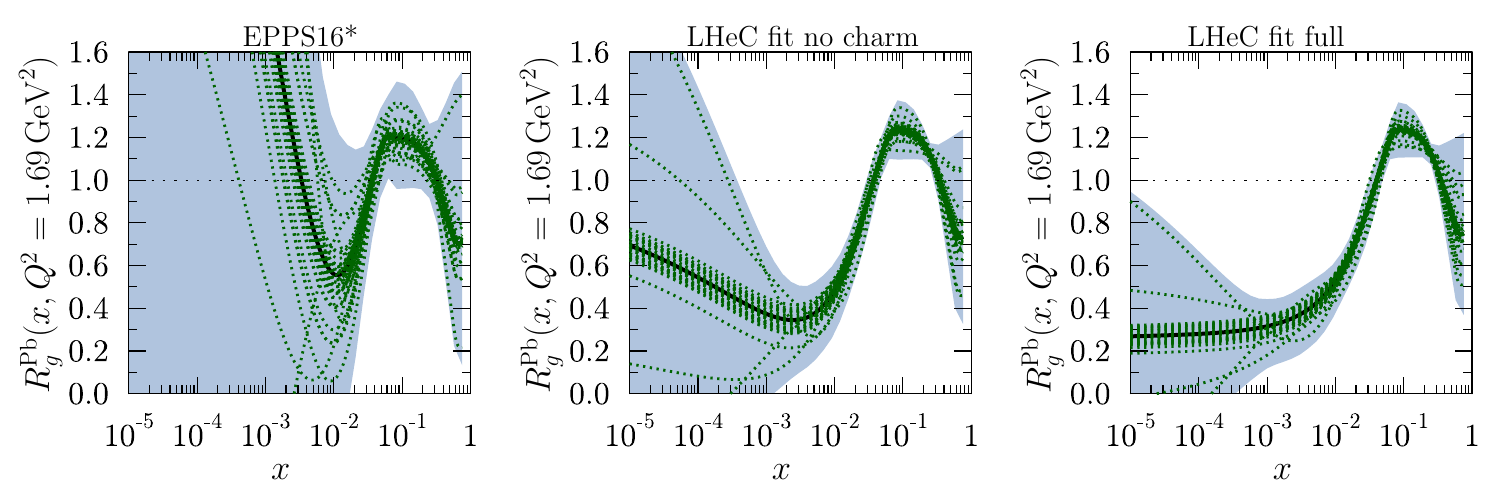}
\caption{Modification to the gluon nuclear parton distribution function at the initial scale~\cite{Agostini:2020fmq}.}
\label{fig:epps16}
\end{figure*}

The nuclear modification factor for the gluon distribution in the EPPS16$^*$ parametrization without the LHeC pseudodata is shown in Fig.~\ref{fig:epps16} (left panel), where the uncertainties become very large in the region where there is no data available. The LHeC impact is analyzed by including the LHeC pseudodata in the fit, with the results shown also in Fig.~\ref{fig:epps16}: the middle panel shows the gluon modification at the initial scale when the total cross section pseudodata is included, and the right panel also includes the charm reduced cross section data. Separate charm reduced cross section data is found to also have a large impact, and the LHeC can accurately constrain the nPDFs down to $x\sim 10^{-4}$ at the initial scale. At larger scales the DGLAP evolution will result in smaller uncertainty bands. The FCC-he would extend the region where the PDFs are determined accurately by even smaller $x$.

The approach in EPPS16 and EPPS16$^*$ is to determine the nuclear PDFs for all nuclei (as a function of $A$). Thanks to the wide kinematical coverage and high precision, with the LHeC or FCC-he it becomes possible to also directly extract the PDF of a single nucleus, e.g. Pb. In this approach there is no need to introduce a dependence on nuclear mass number, and significantly higher precision can be obtained. As reported in~\cite{Agostini:2020fmq}, the LHeC/FCC-he allow for the extraction of the Pb PDF accurately down to $x\sim 10^{-5}$. These accurate measurements will also enable precision tests for the collinear factorization with nuclei.

\section{Diffraction}

In diffractive events $\gamma^* + A \to X + A$ there is no net color charge transfer between the target  $A$ and the produced system $X$, and as such at least two gluons need to be exchanged. Consequently, the cross sections are approximatively proportional to the gluon density squared, and  more sensitive on the small-$x$ structure of nuclei. When considering diffractive processes where all the final state particles can be measured (e.g. exclusive vector meson production) the additional benefit is that the total momentum transfer to the target can be measured. As the momentum transfer is the Fourier conjugate to the impact parameter, these processes provide access to the spatial structure of the target (see e.g. Ref.~\cite{Klein:2019qfb} and references therein).

\begin{figure*}[tb]
\begin{minipage}{0.48\textwidth}
\centering
		\includegraphics[width=\textwidth]{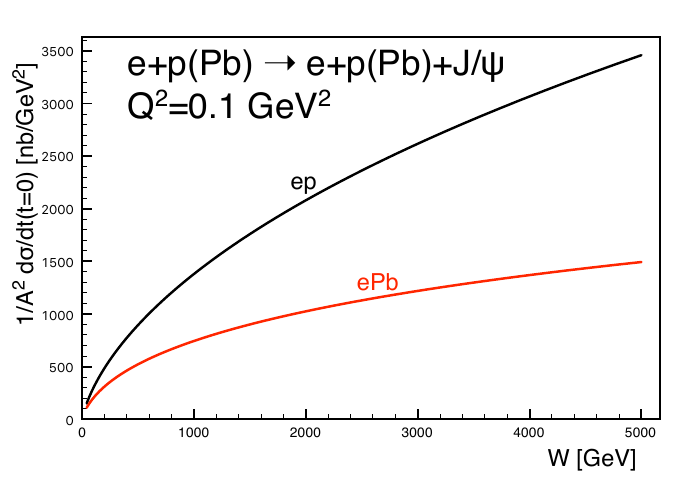} 
				\caption{Exclusive \jpsi production cross section as a function of the photon-nucleon center-of-mass energy $W$~\cite{Agostini:2020fmq}.}
		\label{fig:jpsiWdep}
\end{minipage}
\quad
\centering
\begin{minipage}{0.48\textwidth}
		\includegraphics[width=\textwidth]{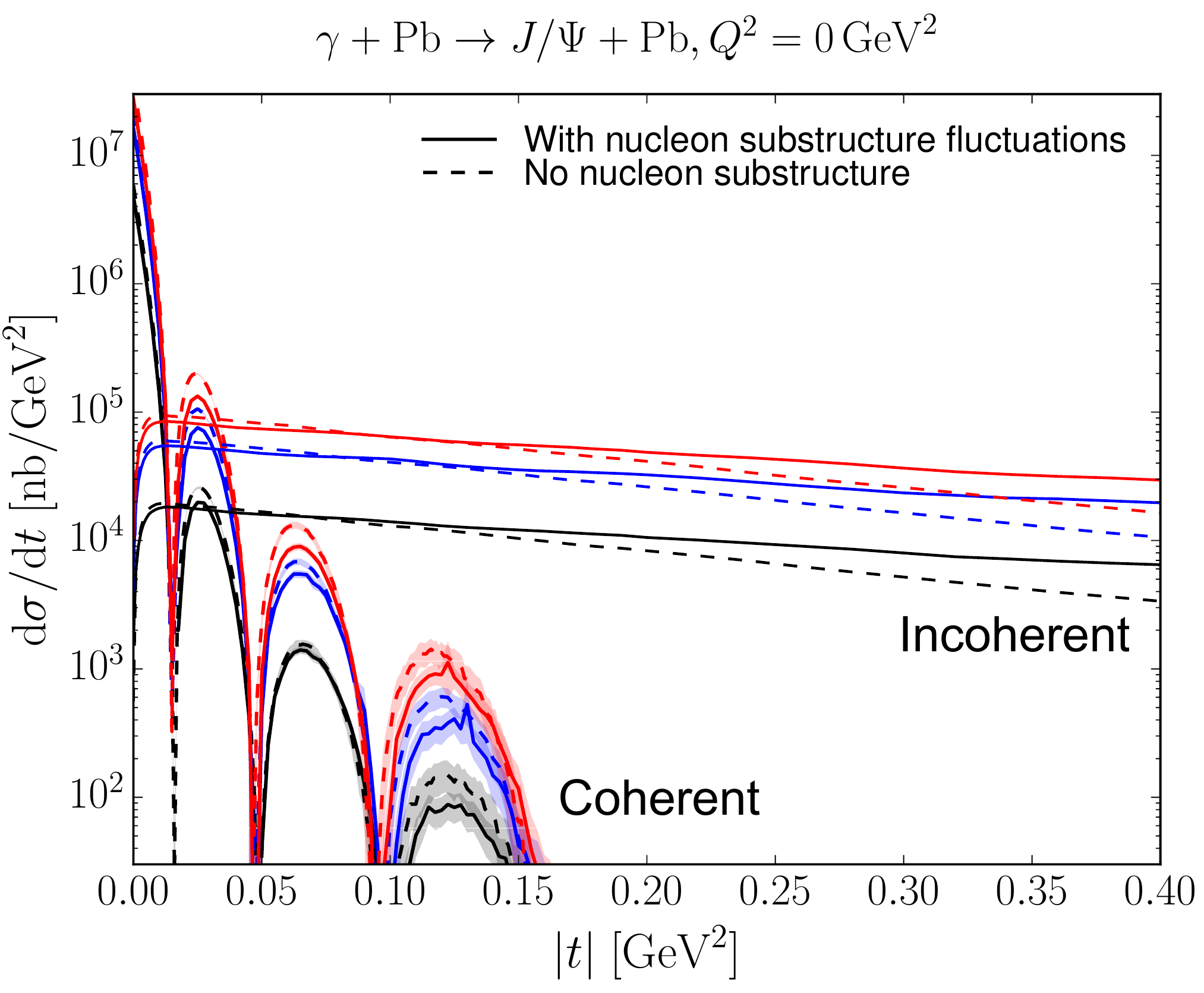} 
						\caption{Coherent and incoherent \jpsi photoproduction cross section as a function of squared momentum transfer at different center-or-mass energies $W=0.1,0.813$ and $W=2.5\,\mathrm{TeV}$ (bottom to top)~\cite{Agostini:2020fmq}.}
		\label{fig:jpsi_spectra}
\end{minipage}
\end{figure*}

 The center-of-mass energy $W$ dependence of the exclusive \jpsi production at zero momentum trasfer $t$ is shown in Fig.~\ref{fig:jpsiWdep} in $e+p$ and $e+\mathrm{Pb}$ collisions. The normalization is chosen such that in the absence of non-linear QCD dynamics (or nuclear effects) the cross sections would be identical. The weaker $W$ dependence in case of a heavy nucleus suggests that non-linear saturation effects become important at high center-of-mass energies. The $\gamma + \mathrm{Pb} \to \jpsim + \mathrm{Pb}$ processes have also been studied in ultra peripheral heavy ion collisions at the LHC (limited to $Q^2=0$), showing a slower increase of the cross section as a function of $A$ than expected without non-linear effects, and measurements are compatible with the predictions based on gluon saturation, see e.g. Refs.~\cite{Lappi:2013am,Khachatryan:2016qhq}. Additional detailed handle on non-linear QCD dynamics can be obtained by studying  exclusive cross sections as a function of $A$ and $Q^2$~\cite{Mantysaari:2017slo} at the LHeC/FCC-he.

If one also considers vector meson production processes where the target can break up (but not exchange net color with the produced meson, the so called incoherent diffraction) one can access not only the average spatial structure of the target, but also the event-by-event fluctuations~\cite{Mantysaari:2016ykx} (see~\cite{Mantysaari:2020axf} for a review). The predicted  coherent (no breakup) and incoherent (with breakup) cross sections with (solid) and without (dashed) proton shape fluctuations constrained by the HERA data in Ref.~\cite{Mantysaari:2016ykx} are shown in Fig.~\ref{fig:jpsi_spectra} in the LHeC/FCC-he kinematics. Studying the coherent and incoherent cross sections at different center-of-mass energies $W$ allows for an extraction of the fluctuating proton and nuclear shapes as a function of Bjorken $x$, recently studied e.g. in Refs.~\cite{Mantysaari:2018zdd,Cepila:2017nef}.

In addition to exclusive particle production, it is possible to study diffractive structure functions, i.e. total diffractive cross section $e+A \to e+X+A$, for example as a function of the invariant mass of the diffractively produced system $M_X^2$. The diffractive structure functions can be related to the diffractive parton distribution functions, that have never been measured for nuclei. An experimental challenge in these measurements with nuclear targets is to distinguish coherent and incoherent events, as incoherent diffraction with nuclear targets starts to dominate  at much smaller momentum transfers $|t|$ than in case of proton targets. In order to separate the two channels (also in exclusive reactions), forward detectors capable of seeing e.g. neutron emission are required.

The large kinematical coverage of the LHeC allows for the measurement of the diffractive structure functions, or the reduced diffractive cross section, over a wide range of $x,Q^2$ and $M_X^2$ (or $\beta \sim 1/M_x^2$). A subset of the generated LHeC pseudodata is shown in Fig.~\ref{fig:diffxs}. The expected high precision of the LHeC is clearly visible, with the uncertainties being dominated by the assumed $5\%$ systematic uncertainty. Consequently, a precise extraction of the nuclear diffractive structure functions is possible~\cite{Agostini:2020fmq}. 

\section{Non linear dynamics at small $x$}

Very high parton densities, or strong color field, accessible at the LHeC/FCC-he allows for numerous opportunities to establish the existence of non-linear QCD dynamics in the nuclear wave function at high energies. Here we briefly mention two possibilities, see Ref.~\cite{Agostini:2020fmq} for a more extended discussion.

The total cross section (structure function) measurements can be used to test the applicability of fixed order perturbative calculations in collinear factorization approach, by studying possible tensions in the DGLAP based fits with different phase space ($x,Q^2$) cuts. Recent analyses suggest that resummation of large logarithms of energy $\alpha_\mathrm{s} \ln 1/x$ might be needed to accurately describe the HERA structure function data~\cite{Ball:2017otu}. High precision and much higher parton densities accessible at the LHeC will make it possible to accurately quantify the role of the non-linear effects in total cross section measurements.

\begin{figure*}[tb]
\begin{minipage}{0.48\textwidth}
\centering
		\includegraphics[width=\textwidth]{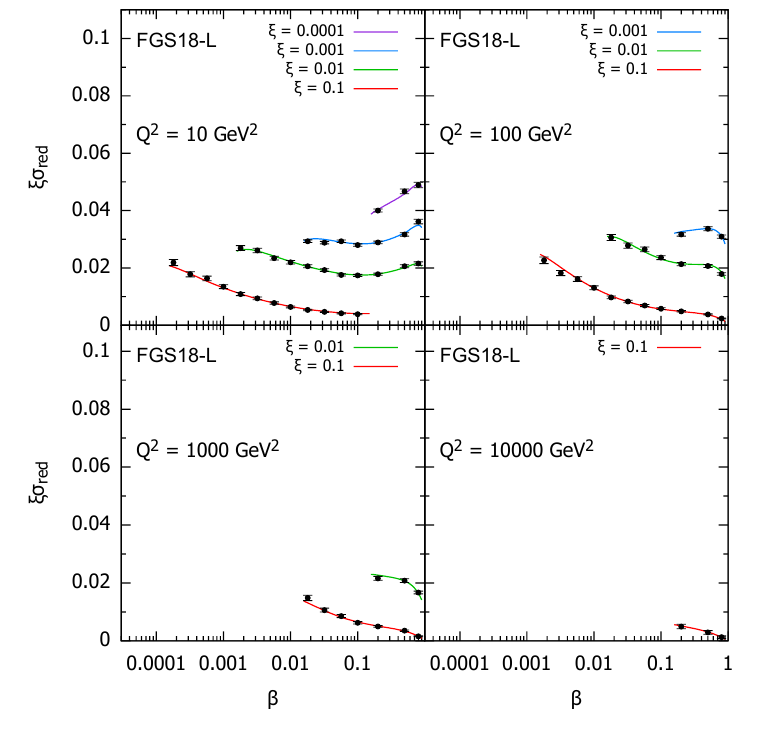} 
				\caption{Simulated diffractive structure function data~\cite{Agostini:2020fmq} in $e+\mathrm{Pb}$ collisions at the LHeC. The small  $\xi$ results are shifted up by $0.01,0.02$ and $0.03$. }
		\label{fig:diffxs}
\end{minipage}
\quad
\centering
\begin{minipage}{0.48\textwidth}
		\includegraphics[width=\textwidth]{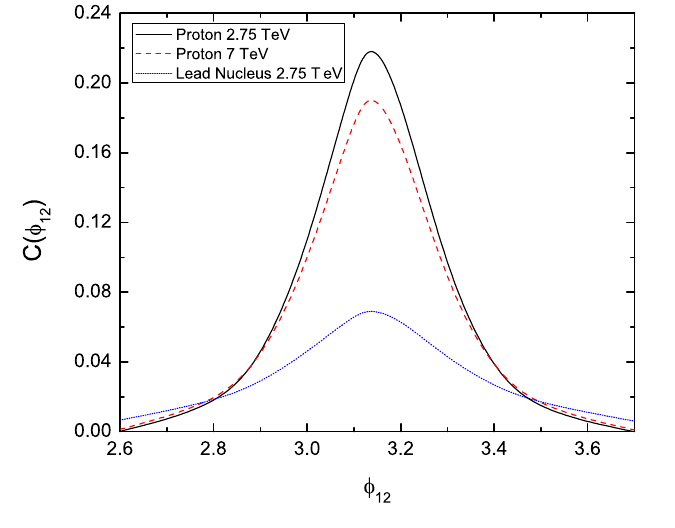} 
						\caption{Azimuthal correlation of two pions in $ep$ and $eA$ collisions at the LHeC~\cite{AbelleiraFernandez:2012cc}. The back-to-back peak is suppressed at higher $\sqrt{s}$ and in heavy nuclei.}
		\label{fig:dijet}
\end{minipage}
\end{figure*}

More differential observables such as two-particle correlations provide access to the more detailed structure of the nuclear wave function at small $x$. One promising process to observe the non-linear QCD dynamics is to look at azimuthal correlations of two hadrons at forward rapidity (electron going direction). In electron-proton collisions, a clear back-to-back correlation is expected which is heavily suppressed in electron-nucleus case. This can be understood to result from the evolution of the saturation scale $Q_s^2$, as the characteristic gluon transverse momentum  scales with $Q_s$. Consequently, at high $Q_s^2$ (Pb target) the gluon transverse momenta are so large that they result in much broader back-to-back peak. The theoretical predictions in the LHeC kinematics are shown in Fig.~\ref{fig:dijet}.

\section{Conclusions}
High energy nuclear DIS experiments at the future LHeC/FCC-he colliders allow for a precise determination of the nuclear structure at high energies. It will also be possible to quantify the role of non-linear QCD dynamics in the very high density region.

\acknowledgments
H.M. is supported by the Academy of Finland project 314764.

\bibliographystyle{h-physrev4mod2}
\bibliography{../../../Dropbox/refs}

\end{document}